\documentclass[12pt,reqno,a4paper,preprint,sort&compress]{elsarticle}

\usepackage{multicol}
\usepackage[english]{babel}
\usepackage{amsmath, amssymb,graphicx,geometry,natbib}
\usepackage{epstopdf}
\usepackage[font={small}]{caption}
\usepackage{float}
\usepackage{color}

\title{Asymptotic expansion approximation for spatial structure arising from directionally biased movement}

\author[uc,tpm]{Michael J. Plank\corref{cor1}}
\ead{michael.plank@canterbury.ac.nz}

\cortext[cor1]{Corresponding author}

\address[uc]{School of Mathematics and Statistics, University of Canterbury, Christchurch 8140, New Zealand}
\address[tpm]{Te P\={u}naha Matatini, Centre of Research Excellence, New Zealand}

\begin{document}

\begin{abstract}
Spatial structure can arise in spatial point process models via a range of mechanisms, including neighbour-dependent directionally biased movement. This spatial structure is neglected by mean-field models, but can have important effects on population dynamics. Spatial moment dynamics are one way to obtain a deterministic approximation of a dynamic spatial point process that retains some information about spatial structure. However, the applicability of this approach is limited by the computational cost of numerically solving spatial moment dynamic equations at a sufficient resolution. We present an asymptotic expansion for the equilibrium solution to the spatial moment dynamics equations in the presence of neighbour-dependent directional bias. We show that the asymptotic expansion provides a highly efficient scheme for obtaining approximate equilibrium solutions to the spatial moment dynamics equations when bias is weak. This scheme will be particularly useful for performing parameter inference on spatial moment models. 
\end{abstract}


\maketitle

\section{Introduction}
Individual-based models incorporating some type of local interaction between neighbouring agents are a widely used tool in mathematical modelling. Populations arising from such models are typically not well mixed, and can exhibit rich spatial structure \citep{mahdi1987spatial,pacala1985neighborhood}. There is a two-way interplay between this spatial structure and population dynamics, which can critically influence population-level outcomes \citep{law2003,murrell03}. Spatial moment dynamics is one way of obtaining a deterministic approximation to a stochastic point process model, while retaining information about spatial structure that mean-field approaches neglect \citep{baker2010correcting,bolker97,dieckmann00}. 

\citet{binny15} introduced a new type of model for interactions among neighbouring agents that was based on neighbour-dependent directionally biased movement. Agents may be biased to move in a direction either towards or away from neighbouring agents. This rapidly generates population-level spatial structure in the form of either clustering if neighbouring agents are attracted towards one another; or separation if neighbouring agents are repelled from one another. This spatial structure can be captured in a spatial moment dynamics approximation \citep{binny16a,binny15}. This offers analytical insights into the key mechanisms driving macroscopic outcomes that cannot be obtained from individual-based models \citep{binny16b}.

Similar neighbour-dependent mechanisms have been modelled by \citet{middleton14}, \citet{matsiaka2017continuum}, \citet{surendran18} and \citet{johnston2019impact}. All of these studies resort to computationally expensive solutions of partial integro-differential equations for the second spatial moment, pair density or comparable correlation function. This limits the usefulness of these approaches, particularly in applications, such as parameter inference, which require the forward problem to be solved a large number of times. Analytical results for lattice-free, many-body systems of interacting agents are more challenging and are largely restricted to limiting cases. For example, \citet{newman2004many} derived a correction to the mean-field diffusion coefficent in a many-body system, using an asymptotic expansion with the chemotactic coupling between cells as a perturbation parameter. \citet{bruna2012diffusion,bruna2012excluded} used matched asymptotic expansions to approximate the pair correlation function and effective diffusivity of hard spheres in the dilute limit. 

In this paper, we derive an asymptotic expansion for the equilibrium spatial moment system, using the bias strength as a perturbation parameter. This parameter plays a similar role to the chemotactic coupling perturbation parameter used by \citet{newman2004many}. However, we use a different model for cell movement based on a neighbour-dependent directional bias mechanism \citep{binny2019living}. This model allows the interactions between neighbouring individuals to be either repulsive \citep{binny16b,browning2018inferring}, attractive \citep{binny16a}, or a combination of repulsion and attraction at differing spatial scales \citep{binny2019living}. We show that the asymptotic expansion offers a highly efficient fixed-point iteration method of obtaining an approximate equilibrium solution for the second spatial moment (equivalent to the pair correlation function). The approximation is valid in the weak bias limit, though additionally requires the average density to be not too large. Nevertheless, even in situations where the solution to the full equilibrium spatial moment dynamics is required, the approximation presented here can still reduce computational overhead by providing an initial condition that is close to equilibrium.

\section{Spatial moment dynamics}
We consider the spatial moment dynamics system of \citet{binny2019living}, which consists of neighbour-dependent directional bias, with no proliferation, mortality or neighbour-dependence in the rate of movement. Individual agents undergo movement events as independent Poisson processes with constant rate $m$ per unit time. The direction of movement depends on the individual's local neighbourhood. A neighbouring agent at displacement $x\in\mathbb{R}^2$ from the motile individual induces a directional bias. The nature of this bias is encoded by a function $v(x)$, referred to as the bias kernel, satisfying $v(x)\to 0$ as $|x|\to\infty$. We assume, without loss of generality, that the spatial variable $x$ has been scaled such that gradients in $v$ are $O(1)$.

\subsection{Dynamics of the second spatial moment}
Since there is no proliferation or mortality, the total population size and hence the first spatial moment (average density, denoted $Z_1$) are fixed. The second spatial moment $Z_2(x,t)$ represents the density of pairs of agents that are separated by a vector $x\in\mathbb{R}^2$ at time $t$ \citep{plank15}. The equation for the dynamics of the second moment is \citep{binny2019living}:
\begin{eqnarray} 
\frac{\partial Z_2(x,t)}{\partial t} = 2m\left(-Z_2(x,t)  + \int Z_2(x+y,t)\mu(y,x+y) dy \right), \label{eq:dZ2dt}
\end{eqnarray}
where 
\begin{equation} \label{eq:bivariate_pdf}
\mu(y,z) = f(y) \frac{\exp\left(|\eta(z)|\cos\left(\arg(y)-\arg\left(\eta(z)\right)\right)\right)}{I_0\left(|\eta(z)|\right)} 
\end{equation}
is the bivariate probability density function (PDF) for moving by $y$, conditional on the presence of another agent at displacement $z$, and
\begin{equation}\label{eq:bias_vector}
\eta(z) = \beta\left( \nabla v(z) +  \int \nabla v(y) \frac{Z_3(z,y)}{Z_2(z)}dy \right)  
\end{equation}
is the expected net bias vector, conditional on the presence of another agent at displacement $z$. Here, $I_0$ denotes the modified Bessel function of the first kind and order zero. All integrals are over $\mathbb{R}^2$. The factor of $2$ in Eq. (\ref{eq:dZ2dt}) accounts for movement of the two individuals in the pair \citep{binny2019living}, and makes use of the symmetries $Z_2(x,t)=Z_2(-x,t)$ and $\mu(x,y)=\mu(-x,-y)$ .

The bivariate PDF defined by Eq. (\ref{eq:bivariate_pdf}) consists of a function $f(y)$, which depends only on the distance moved $|y|$, and a von Mises distribution for the direction of movement \citep{binny16a}. The concentration and mean direction of the von Mises distribution are set to the magnitude and direction respectively of the expected bias vector $\eta(z)$. Since $\mu(y,z)$ is a probability density function with respect to $y$ for any fixed value of $z$, it is required that $\int \mu(y,z) dy=1$. This is satisfied provided that $\int f(y) dy=1$.

The net bias vector in Eq. (\ref{eq:bias_vector}) is the expected value of the bias vector of an agent in a pair with another agent at displacement $z$. The first term on the right-hand side of Eq.  (\ref{eq:bias_vector}) is the direct contribution, $\nabla v(z)$, from the other agent in the pair and the second term is the contribution, $\nabla v(y)$, from a third agent at displacement $y$, weighted by the probability of an agent being located at displacement $y$, integrated over $y$ (see \citep{binny2019living} for details).

\subsection{Moment closure approximation}
Eqs. (\ref{eq:dZ2dt})--(\ref{eq:bias_vector}) are not a closed system because they depend on the third moment $Z_3$. In keeping with previous work, we use a moment closure scheme that approximates the third moment $Z_3$ in terms of the first and second moments. Various moment closure schemes have been proposed \citep{murrell04}. We use the well-known \citet{kirkwood35} superposition approximation, which is the most analytically tractable, as well as having comparable accuracy with other closure schemes for the third moment in a variety of scenarios \citep{omelyan2019spatial},
\begin{equation}\label{eq:ksa}
Z_3(x,y,t) = \frac{Z_2(x,t)Z_2(y,t)Z_2(y-x,t)}{Z_1^3}.
\end{equation}
Under this moment closure scheme, Eq. (\ref{eq:bias_vector}) becomes
\begin{equation}\label{eq:bias_vector_closed}
\eta(z) = \beta\left( \nabla v(z) +   \frac{1}{Z_1^3} \int \nabla v(y) Z_2(y)Z_2(z-y) dy \right). 
\end{equation}
Together with an initial condition for $Z_2(x,0)$ and the boundary condition $Z_2(x,t)\to Z_1^2$ as $|x|\to\infty$, this defines a closed dynamical system for $Z_2(x,t)$.

\section{Equilibrium solution for the second moment}
Writing an equilibrium solution to Eq. (\ref{eq:dZ2dt}) as $Z_2(x)=Z_1^2 \left(1+h(x)\right)$, we have 
\begin{equation} \label{eq:h_ss}
0 = -1-h(x)  + \int \left(1+h(x+y)\right)\mu(y,x+y) dy.
\end{equation}
The cosine function in Eq. (\ref{eq:bivariate_pdf}) can be recast as a scalar product of the movement vector $y$ and the expected bias vector $\eta(z)$ to give
\begin{equation} \label{eq:bivariate_sp}
\mu(y,z) = f(y) \frac{\exp\left(\frac{y.\eta(z) }{|y|} \right)}{I_0\left(|\eta(z)|\right)}. 
\end{equation}
Noting that $\int q(x) \nabla v(x) dx=0$ for any even function $q$, the expected bias vector in Eq. (\ref{eq:bias_vector_closed}) can be written in convolution form:
\begin{equation} \label{eq:bias_vector_conv}
\eta = \beta\left( \nabla v +  Z_1 \left( (1+h)\nabla v\right) * h\right).
\end{equation}
Eqs. (\ref{eq:h_ss})--(\ref{eq:bias_vector_conv}) form a compact system of equations for the equilibrium solution $h(x)$.

\subsection{Asymptotic expansion}
Now expanding Eq. (\ref{eq:bivariate_sp}) as a Taylor series in $\eta$ neglecting terms of order $|\eta|^3$ and higher gives:
\begin{equation} \label{eq:bivariate_pdf_taylor}
\mu(y,x+y) = f(y) \left(1 + \frac{y.\eta(x+y) }{|y|} + \frac{(y.\eta(x+y))^2 }{2|y|^2} - \frac{|\eta(x+y)|^2}{4} \right).
\end{equation}
Substituting this into Eq. (\ref{eq:h_ss})  gives
\begin{eqnarray} 
0 &=& -h(x)  + \int f(y)h(x+y) dy \label{eq:h_intermediate} \\
&& + \int f(y) \left(1+h(x+y)\right)\left(\frac{y.\eta(x+y) }{|y|} + \frac{(y.\eta(x+y))^2 }{2|y|^2} - \frac{|\eta(x+y)|^2}{4} \right) dy. \nonumber
\end{eqnarray}
We now make use of convolution notation, which will allow terms in the asymptotic expansion for $h(x)$ (see below) to be written compactly. For two real-valued functions on $\mathbb{R}^2$, $\phi_1$ and $\phi_2$, we use the standard notation $(\phi_1 * \phi_2)(x) = \int \phi_1(z)\phi_2(x-z) dz$. Making a change of integration variable $z=-y$ in Eq. (\ref{eq:h_intermediate}), using the fact that $f$ is even, and omitting the $x$ argument for convenience gives
\begin{eqnarray} \label{eq:convolution}
h &=& (f*h) - \sum_{i\in\left\{1,2\right\}} \left(\frac{fx_i}{|x|}\right)  * \left((1+h)\eta_i)\right) + \frac{1}{2} \sum_{i,j\in\left\{1,2\right\}} \left(\frac{fx_i x_j}{|x|^2}\right)*\left((1+h)\eta_i \eta_j\right) \nonumber \\
&& - \frac{1}{4}\left( f*((1+h)|\eta|^2)\right).
\end{eqnarray}
We now write $h(x)$ as an asymptotic expansion in the bias strength parameter $\beta$: 
\begin{equation} \label{eq:asymptotic_expansion}
h(x)= \sum_{k=0}^\infty \beta^k h_k(x).
\end{equation}
When there is no directional bias ($\beta=0$), $\mu(y,z)$ is equal to $f(y)$ and hence $h(x)=0$ is a solution to Eq. (\ref{eq:h_ss}). This implies that the leading order term $h_0(x)$ in Eq. (\ref{eq:asymptotic_expansion}) is zero. Under the asymptotic expansion for $h(x)$, Eq. (\ref{eq:bias_vector_conv}) for the expected net bias vector becomes
\begin{equation} \label{eq:bias_vector_expansion}
\eta = \beta \nabla v +  \beta^2 Z_1 \left( \nabla v * h_1\right) + O(\beta^3),
\end{equation}
provided the average density $Z_1$ is $O(1)$. Substituting this expression for $\eta$ into Eq. (\ref{eq:convolution}) and equating terms of order $\beta^1$ and $\beta^2$ respectively gives
\begin{eqnarray}
h_1 &=& (f*h_1) - \sum_{i\in\left\{1,2\right\}} \left(\frac{fx_i}{|x|}\right)  * \frac{\partial v}{\partial x_i},    \label{eq:iterative1} \\
h_2 &=& (f*h_2) - \sum_{i\in\left\{1,2\right\}} \left(\frac{fx_i}{|x|}\right)  * \left(h_1\frac{\partial v}{\partial x_i} + Z_1h_1*\frac{\partial v}{\partial x_i}\right)  \nonumber \\ 
&& + \frac{1}{2} \sum_{i,j\in\left\{1,2\right\}} \left(\frac{fx_i x_j}{|x|^2}\right)*\left(\frac{\partial v}{\partial x_i} \frac{\partial v}{\partial x_j} \right) 
- \frac{1}{4} \left( f*(|\nabla v|^2)\right). \label{eq:iterative2} 
\end{eqnarray}
In principle, equating terms of order $\beta^3$ gives a similar equation for $h_3$ (this would require terms of order $|\eta|^3$ to be retained in Eq. (\ref{eq:bivariate_pdf_taylor})), and so on. This results in a hierarchy of equations of the form 
\begin{equation}\label{eq:iterative_general}
h_k = (f*h_k) + C_k\left(h_1,\ldots,h_{k-1}\right),
\end{equation}
where $C_k$ is some function of the lower-order terms in the asymptotic expansion up to $h_{k-1}$.  Solutions to these convolution equations are not readily available in closed form. Nevertheless, they offers an efficient fixed-point iteration method for calculating terms in the asymptotic expansion for $h(x)$, by first iterating Eq. (\ref{eq:iterative1}) until $h_1$ converges, then iterating Eq. (\ref{eq:iterative2}) until $h_2$ converges, and so on. Although the number of terms in Eq. (\ref{eq:iterative_general}) increases with $k$, each iteration only requires computation of the convolution $(f*h_k)$. Hence, the computational requirements for calculating each term in the asymptotic expansion are similar.

\begin{table}
    \centering
    \begin{tabular}{p{8cm}ll}
    \hline
         {\bf Kernels} && \\
         Movement distribution & $f(x)$ &  $Ce^{-\frac{(x-x_m)^2}{2\sigma_m^2} }$ \\
         Interaction kernel 1 (short-range repulsion) & $v(x)$ & $e^{-\frac{x^2}{2\sigma_{b}^2}}$ \\
         Interaction kernel 2  (short-range repulsion, medium-range attraction) & $v(x)$ & $e^{-\frac{x^2}{2\sigma_{b}^2}} - k_2e^{-\frac{(x-x_b)^2}{2\sigma_{b}^2}}$ \\\hline
         {\bf Parameters} && \\
         Average agent density & $Z_1$ &  1\\
         Bias strength & $\beta$ & varied \\
         Characteristic step length & $x_m$ & $1.25$ \\
         Step length standard deviation & $\sigma_m$ & $0.5$ \\
         Characteristic repulsive interaction length scale & $\sigma_{b}$ & $1$  \\
         Characteristic attractive interaction length scale & $x_{b}$ & $3$  \\
         Strength of attraction relative to repulsion & $k_2$ & $0.5$ \\
         Size of computational domain & $x_\mathrm{max}$ & $9$ \\ 
         Mesh point spacing & $\delta x$ & $0.1$ \\
         \hline
    \end{tabular}
    \caption{Model kernels and parameter values. The movement distribution $f(x)$ is truncated at $x=x_m+3\sigma_m$ and $C$ is a normalisation constant chosen such that the distribution integrates to $1$ over the computational domain. }
    \label{tab:1}
\end{table}

\subsection{Accuracy of asymptotic expansion}
To test the accuracy of the asymptotic expansion, we compute the first and second order terms $h_1$ and $h_2$, and compare to a full solution found by performing fixed-point iteration directly on the equilibrium equation Eq. (\ref{eq:h_ss}). Note that $1+h(x)$ is equivalent to the pair correlation function at equilibrium. Hence, negative values of $h(x)$ correspond to a regular spatial structure, whereas positive values of $h(x)$ correspond to a clustered spatial structure. Parameter values, movement distribution and interaction kernel are shown in Table 1.

\begin{figure}
    \centering
    \includegraphics[width=1.1\textwidth]{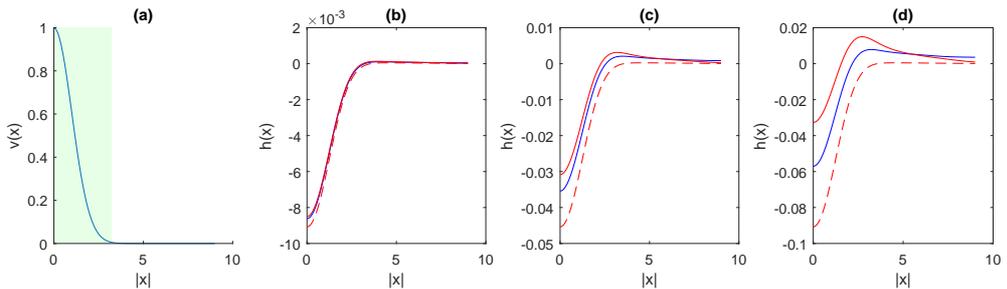} 
    \caption{Results for interaction kernel 1: (a) interaction kernel $v(x)$ generates repulsion in the range where $dv/dx<0$ (green); (b)-(d) comparison of the full solution $h(x)$ to Eq. (\ref{eq:h_ss}) (solid blue) with the asymptotic approximation up to $O(\beta^1)$ (dashed red) and $O(\beta^2)$ (solid red), for bias strength (a) $\beta=0.01$; (b) $\beta=0.05$; (c) $\beta=1$. Other parameters as shown in Table 1.  }
    \label{fig:repulse}
\end{figure}

Figure \ref{fig:repulse} shows the results under short-range repulsion between neighbouring agents (Fig. \ref{fig:repulse}a, interaction kernel 1 in Table 1). The graphs compare the full solution $h(x)$ (blue), the first order approximation $\beta h_1(x)$ (dashed red), and the second order approximation $\beta h_1(x)+\beta^2 h_2(x)$ (solid red). Since $h(x)$ is axisymmetric, a cross-section of $h(x)$ is plotted against $|x|$. As expected, the spatial structure is regular ($h(x)<0$ for small $|x|$), meaning that there are fewer pairs of agents close to one another than there would be under a Poisson point process of the same average density. For weak bias ($\beta=0.01$), the asymptotic expansion including terms up to $O(\beta^2)$ provides an excellent approximation to the full solution (Fig. \ref{fig:repulse}a). As bias strength increases, the spatial structure becomes stronger (larger $|h(x)|$), and the accuracy of the asymptotic expansion begins to decrease (Fig. \ref{fig:repulse}b-c), although it is still a reasonable approximation at $\beta=0.05$.

\begin{figure}
    \centering
    \includegraphics[width=1.1\textwidth]{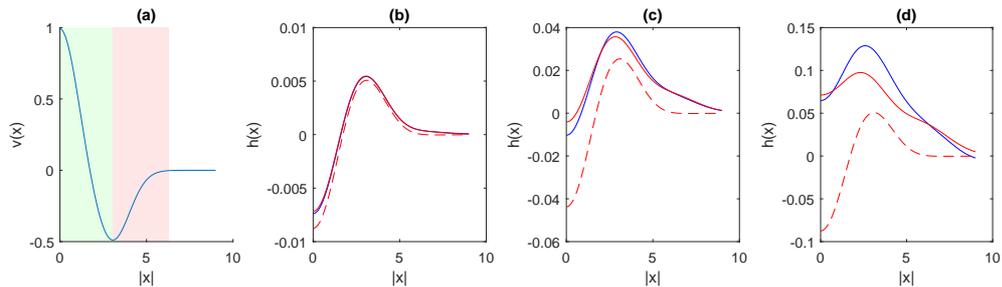}
    \caption{Results for interaction kernel 2: (a) interaction kernel $v(x)$ generates repulsion in the range where $dv/dx<0$ (green) and attraction in the range where $dv/dx>0$ (red); (b)-(d) comparison of the full solution $h(x)$ to Eq. (\ref{eq:h_ss}) (solid blue) with the asymptotic approximation up to $O(\beta^1)$ (dashed red) and $O(\beta^2)$ (solid red), for bias strength (a) $\beta=0.01$; (b) $\beta=0.05$; (c) $\beta=1$. Other parameters as shown in Table 1. }
    \label{fig:repulse_attract}
\end{figure}

Figure \ref{fig:repulse_attract} shows the results under a combination of short-range repulsion and medium-range attraction (Fig. \ref{fig:repulse_attract}a, interaction kernel 2 in Table 1). This leads to a regular structure ($h(x)<0$) at short spatial scales and clustered structure ($h(x)>0$) at medium scales, representing individuals that form groups, but are separated from their immediate neighbours \citep{binny2019living}. Again the asymptotic expansion provides a good approximation to the full solution up to $\beta=0.05$, and starts to lose accuracy for larger values of $\beta$. The average density is set at $Z_1=1$ for the results in Figures \ref{fig:repulse} and \ref{fig:repulse_attract}. The accuracy of the approximation decreases as the value of $Z_1$ increases, though is less sensitive to $Z_1$ than it is to $\beta$.

The asymptotic expansion is several orders of magnitude more efficient to compute, taking approximately 1 s to calculate the linear approximation in Fig. \ref{fig:repulse}(a) and a further 1 -- 2 s to calculate the quadratic term, compared to over $5000$ s to calculate the solution of the full equilibrium equation (\ref{eq:h_ss}). Using the asymptotic expansion as the initial condition for fixed point iteration of Eq. (\ref{eq:h_ss}) instead of a naive initial condition greatly reduces the computation time for the full solution (see Table 2). This shows that, even if the more accurate solution of the full equilibrium equation is required, this can be obtained more efficiently by first computing the asymptotic expansion.  
All computations were done using a $181\times 181$ numerical grid (see Table 1 for numerical parameters) on an Intel(R) Core(TM) $1.70$ GHz CPU with 4 cores and $16$ GB of RAM. 
Matlab's {\em trapz} function was used to compute the integral in Eq. (\ref{eq:h_ss}) and {\em conv2} function was used to compute the convolutions in Eqs. (\ref{eq:bias_vector_conv}), (\ref{eq:iterative1}) and (\ref{eq:iterative2}).

\begin{table}
    \centering
    \begin{tabular}{llll}
    Solution type & Initial condition & \multicolumn{2}{l}{Computation time (s)} \\
     &&  Kernel 1 & Kernel 2 \\
    \hline
    Linear        & $h_1(x)=0$ &  $1.0$& $1.2$ \\
    Quadratic     & $h_2(x)=0$ & $1.3$ & $1.6$ \\
    Full & $h(x)=0$ & $5445$ & $6557 $\\
    Full  & $h(x)=\beta h_1(x)+\beta^2 h_2(x) $ & $1441$ & $337 $  \\
    \end{tabular}
    \caption{Comparison of computation times required for different solution types and the two interaction kernels with $\beta=0.01$: `linear' is the time taken to compute the first term in the asymptotic expansion; `quadratic' is the additional time taken to compute the first term in the asymptotic expansion; `full' is the time taken to compute the solution to the full equilibrium equation (\ref{eq:h_ss}) either using a naive initial condition (IC) specified by $h(x)=0$, or using the quadratic approximation as the initial condition. All solutions are calculated via fixed point iteration until the relative change in the solution between consecutive iterations is less than $\epsilon=10^{-2}$ in norm.}
    \label{tab:2}
\end{table}

\section{Discussion}
Individual-based models that include movement with neighbour-dependent directional bias has been shown to be an effective way to model spatially structured populations, including cell populations \citep{binny16b}, and individuals living in groups \citep{binny2019living}. However, performing individual-based stochastic simulations alone is computationally expensive and offers limited insight into the key mechanisms. Spatial moment dynamics is one way to obtain a deterministic approximation of the individual-based stochastic process that retains some information about spatial structure, which mean-field models neglect \citep{bolker97,dieckmann00}. Spatial moment dynamics has given analytical insights into the the relationship between individual-level mechansisms and population-scale outcomes that individual-based models alone cannot \citep{binny2019living,law2003,murrell03}. However, the effectiveness of spatial moment dynamics as an approximation to individual-based models of movement with neighbour-dependent bias is constrained by the computational demands of solving for the second moment \citep{binny16a,binny16b}. 

In this paper, we have shown that an asymptotic expansion for the second moment in terms of the bias strength parameter is a highly efficient method of obtaining an approximate solution for the second moment at equilibrium. The asymptotic expansion can be computed via fixed point iteration at least $1000$ times faster than the full equilibrium solution. Although formally valid in the weak bias limit, the asymptotic expansion provides a reasonable approximation even at moderate levels of bias that generate significant spatial structure. Furthermore, the asymptotic expansion can be obtained cheaply and used as an initial condition for the full equilibrium solution, substantially reducing the required convergence time. We have explicitly computed the first and second order terms in the asymptotic expansion, but it would be possible in principle to compute higher-order terms to increase accuracy. Because of the structure of the terms in the asymptotic expansions, each fixed-point iteration only requires computation of a single convolution, and so the computational overhead is approximately linear in the number of terms required. This therefore offers a highly efficient route to increasing accuracy. 

We have focused on a simple form the model which consists only of movement with neighbour-dependent directional bias, but is nevertheless capable of generating rich spatial structure \citep{binny2019living}. More general models include other mechanisms, such as proliferation, dispersal and mortality \citep{binny16b}. Frequently in such models, movement events occur far more frequently than proliferation or mortality events \citep{baker2010correcting}. At a given average density, the population spatial structure is therefore likely to be in pseudo-steady state as determined by the movement component. We have examined a translationally invariant form of the model, in which the ensemble average agent density is independent of spatial location. Equations for spatial moment dynamics have been derived in the more general, translationally dependent case \citep{binny16a,omelyan2019spatially,plank15}. However, due to prohibitive computational costs, most numerical solutions have been restricted to the translationally invariant case, except in special cases \citep{bolker2003combining,lewis2000spread,lewis2000modeling,omelyan2019spatially}. The technique presented in this paper is only applicable to equilibrium solutions, whereas for translationally dependent problems the transient solution is typically of more interest. Nevertheless, the asymptotic expansion technique may still prove useful as a fast method of generating initial conditions for a more computationally expensive time-stepping routine. 

Computationally efficient methods are particularly important in applications, such as parameter inference, where the forward problem (computation of the equilibrium solution for given parameter values) needs to be solved multiple times. Parameter inference has been performed on a model with neighbour-dependent directional bias using approximate Bayesian computation (ABC) on data from cell proliferation assays \citep{browning2018inferring}. This study used ABC rejection sampling on the individual-based model, which is computationally expensive. Although there are more efficient ABC algorithms, such as sequential Monte Carlo \citep{sisson2007sequential}, these still require the forward problem to be solved many times. Inference procedures such as these could be made more efficient by using a fast method of obtaining an approximate solution, which can be used to rapidly reject samples from regions of parameter space in which the posterior density is extremely low. Fast approximate methods can also be used as part of a hierarchical Bayesian framework \citep{maclaren2017hierarchical}. 

The fast approximate method presented here also opens up the possibility of performing likelihood-based Bayesian inference \citep{warne2019simulation} directly on the spatial moment dynamics approximation, which has not yet been attempted. This would require a likelihood function for observed data on individual locations. This could be approximated by using a Palm intensity function \citep{illian2008statistical,tanaka2008parameter} to express the likelihood in terms of the second moment evaluated over all pairs of individuals in the data.

\subsection*{Acknowledgements}
The author acknowledges financial support from Te P\={u}naha Matatini.

\bibliography{bibfile}

\end{document}